\magnification\magstep2
\overfullrule 0 pt
\baselineskip=18pt

\centerline{\bf Excited diskrete spectrum states}
\centerline{\bf wave functions of quantum integrable}
\centerline{\bf N-particle systems in an external field}

\vskip 1 in
\centerline{\bf D.V.Meshcheryakov, V.B.Tverskoy}
\vskip 0.5 in
\centerline{\it Physics Department, Moscow State University,}
\centerline{\it Leninskie gory, 119899, Moscow, Russia}

\vskip 1 in
\centerline{ Abstract }
 A problem of constructing excited state swave functions of the
discrete spectrum of completely integrable quantum systems is
considered.  Recurrence relations defining wave functions up to the
normalizing constant are obtained.
\vskip 1 in

\centerline{\bf 1. Introduction}

Explicit solutions of classical and quantum completely
integrable systems still attract much attention[1-5].

{\ \ \ }By now a lot of completely integrable systems with $N$
degrees of freedom have been discovered[3-5].  However for few of
such systems explicit solutions have been found. Some methods use
additional constants of motion to reduce and simlify corresponding
eigenvalue problems.

  In [6] we proposed a method of finding descrete spectrum energies
and wave functions. This method doesn't use the fact of existence
of additional constants of motion. In [7] we applied this method to
the quantum systems with the following Hamilton function
$$ H= \sum_{i=1}^{N} \left({p_i^2 \over 2} + W(x_i) \right)
+ \alpha(\alpha+1)\sum_{i>j}^N[V(x_i -x_j)+\epsilon V(x_i +x_j)
]\eqno(1) $$
with pair interaction potentials
$$ V(x)={ 1 \over {sh^2x}},{~~~}\epsilon =0,1$$
External field for $\epsilon =1 $ has the form
$$W(x)= {{\mu(\mu-1)}\over 2 }{ 1 \over {sh^2x}}-
{{\lambda(\lambda+1)}\over 2 }{ 1 \over {ch^2x}},{~~~} \lambda >0
\eqno(2a) $$
whereas for $\epsilon =0 $ it takes the form
$$ W(x)= 2A^2(e^{4x}-2e^{2x}) \eqno(2b) $$
The first step of the method proposed in [6] is the
transformation of variables $t_j=(chx_j)^{-2},{~}j=1,...,N$, which
enables us to reduce the order of sigularities in Schrodinger
equation $H\psi=E\psi$. Then we perform furhter transformation to
variables symmetrical in $t_j,{~}j=1,...,N$
$$ a_1=\prod_{j=1}^N t_j, {~~~} a_l={{\hat D^{l-1}} \over
{(l-1)!}}a_1, {~~~} \hat D= \sum_{j=1}^N {\partial \over {\partial
t_j}},{~}l=1,...,N $$
which enables us to convert Schrodinger equation into the form
convenient for solving it explicitly
$$ [H_1(\beta) +
H_2(\beta)] \kappa=[E-\epsilon(\beta)] \kappa $$
$$  H_1(\beta) = 2 \sum_{l,m=1}^N[\sum_{\tau
=1}^{l-1}\sum_{\nu=0}^1((l+m-2\tau)-\nu)a_{\tau}a_{l+m-\tau - \nu} -
(N-m+2)a_{m-1}a_l$$
$$-(N-l+2)a_ma_{l-1}-(N+m+1)a_la_m]{\partial^2 \over
{\partial a_l \partial a_m}} \eqno(3)$$
$$- \sum_{l=1}^N[a_{l-1}(N-l+2)(4\beta +
2\mu +3+ 2\alpha(N+l-3)) + 2a_l (n-l+1)(2\beta + 1 + \alpha (N+l-2)]
{\partial \over \partial a_l},$$
$$ H_2(\beta)=a_N( 2 \sum_{l,m=1}^Na_l a_m {{\partial^2} \over
{\partial a_l \partial a_m}}$$
$$ + \sum_{l=1}^N a_l{\partial
\over {\partial a_l}}(4 \beta + 2 \mu +3 +4 \alpha (N-1) +
p(\beta))$$
Here parameter $\beta$  and $\epsilon(\beta) ,
p(\beta)$ are to be determined from normalizing
condition. Wave functions are polinomials in symmetrical variables
${a_j}$    :
$$ \psi (x_1,...,x_N) = \kappa(a_1,...,a_N) \prod_{j=1}^N
z_j^\mu(1-z_j^2)^\beta \prod_{j>k}(z_j^{1+\epsilon}-z_k^{1+\epsilon})
$$ где $z_j=thx_j $ and
$$ \kappa(a_1,...,a_N)= \sum_{p=0}^{n}\sum_{0 \le j_1,...,j_N \le N}
c_{j_1,...,j_N} a_1^{j_1}...a_N^{j_N} \eqno(4)$$
Energy levels are determined by a set of $N$ integers ${j_1,...,j_N},
0 \le j_1,...,j_N \le n, j_1+...+j_N=p \le n$ , where $n$ --
is the highest powerof the polinomial $\kappa$.  One can conclude
from the normalizing condition that the upper boundary for $n$
is
$$ n <{{\lambda - \mu} \over 2}-\alpha (N-1)
,{~~~}\epsilon =1 $$ $$ n < A-\alpha (N-1) -{1 \over 2}
,{~~~}\epsilon =0 $$
Energy levels are the following
$$ -E_{j_1,...,j_N} = - \epsilon(\beta) - 4 \sum_{l>m}^Nlj_lj_m +
2(N+1)p(\beta)(p(\beta)+ \lambda - \mu -2n- \alpha N)$$
$$+ 2 \sum_{l=1}^N lj_l[\alpha (2N+1) + 2n + \mu - \lambda -j_l -
\alpha l] \eqno(5)$$
In  [7] we only considered the simplest case $n=1$ . Here we find
wave functions of excited levels with $n=2$.

\centerline{\bf 2. Poschl-Teller potential}

Expression (2а) is a generalized Poschl-Teller potential.
Having demanded wave functions to have the following form
$$ \kappa(a_1,...,a_N)= \sum_{l=0}^{N}c_l a_l +c_{N+1} +
    \sum_{l \le k}^N b_{lk} a_l a_k
\eqno(6)$$
and taking into account the normilizing condition we come to
$$ \beta =  {{\lambda - \mu} \over 2}-\alpha (N-1) - 2 $$
$$ p(\beta)= 1-2 \lambda \eqno(7)$$
$$ \epsilon(\beta)=-{N \over 2}[{\alpha^2 \over 3}(N^2-1)
+(\lambda -\mu - \alpha (N-1) - 4)^2]$$
Substititing (5), (6) and (7) into (3), we come to a system
of ${{N(N+1)} \over 2}$ linear algebraic equations with
zero determinant. Unlike the case $n=1$, there is a contribution
from the first term in
$H_1(\beta)$   with second order derivatives in $a_j$.
This leads to coefficients
$c_l,{~}b_{lk}$ being dependent on all non-zero coefficients
 $c_m,{~}m>l,
{~~}b_{mn},{~}m>l,n>k$.  Performing induction one can obtain
recurrence relations for coefficients in (6).  As a result we come
to 3 different sets of recurrent relations for different meanings of
quantum numbers $j_1,...,j_N$.

1. In this case  $j_l = 2 \delta_{ls},
{~}s=1,...,N,{~}p=2;{~}s'=2s-N-1,{~}n'=2s-j, {~}K=(N-s+1)(\lambda
-\mu -2 - \alpha (N-s)); {~}b_{ss}=d_s$ , where $d_s$ is an arbitrary
constant to be determined from the normilizing condition, and
$$ b_{jl}=0,{~}j+l>2s;{~~~}c_j=0,{~}j>s'$$
$$ b_{jn'}= {{4(n'-j)} \over F_{j00}} \sum_{\sigma=j+1}^s
b_{\sigma n'- \sigma +j} \eqno(8)$$
where $F_{jkl}$ are defined below.

2. In this case $j_l =  \delta_{ls} + \delta_{lN},
{~}s=1,...,N-1,{~}p=2;{~}s'=s-1,{~}n'=N, {~}K= \mu -\lambda
+1; {~}b_{sN}=d_s$ , where $d_s$ is an arbitrary
constant to be determined from the normilizing condition, and
$$ b_{jl}=0,{~}j+l>s+N;{~~~}c_j=0,{~}j>s'$$

Other relations have the form
$$ b_{sN}={{1-2 \lambda} \over Q_s} d_s$$
$$b_{s-m N}= {1 \over Q_{s-m}}\{ -2[(N-s+m+1)(2 \lambda -3 -2
\alpha(N-s+m)+1]b_{s-m+1N} $$
$$+4(N-s+m) \sum_{k=1}^{m}\sum_{\tau =0}^1
b_{s-m+k N-k+\tau} + (1-2 \lambda) c_{s-m}\},{~~}m=1,...,s-1 $$
In this relation summation is performed under the condition
 $2k \le N-s+m + \tau$.

It must be stressed here that Hamiltonian (1) is
self-conjugated if and oly if  $(2\lambda
-1)>10+4 \alpha(N-1)$, i.e. the coefficients $c_j$
and $b_{jk}$ are not independent.

3. In this case $j_l =  \delta_{ls} + \delta_{lN-m},
{~}s=1,...,N-2,{~}m=1,...,N-s+1{~},{~}b_{sN-m}=d_{sN-m},{~}s'=s-m-1,{~}
n'=N+s-m-j, {~} $ , where $d_s$ is an arbitrary
constant, and
$$ b_{jl}=0,{~}j+l>s';{~~~}c_j=0,{~}j+l>N+s-m$$
Coefficients $b_{jn'}$ are determined by (8).
Suppose
$$ R_{jkl}=2^l \prod_{\tau=1}^l[(s-j+k- \tau)( \lambda - \mu -3 -
\alpha (2N-s-j+k- \tau))-K],$$
$$ F_{jkl} = 2^{k-l} \prod_{\tau=0}^{k-l} [(s-j)( \lambda - \mu -3 -
\alpha (2N-s-j+1))$$
$$+(N-n'+k- \tau+1)( \lambda - \mu -1 -
\alpha (N-n'+k- \tau))-K],$$
$$ S_j= R_{j11}+ N-j+1,$$
$$ Q_j= R_{j11}+ 2(N-s+1)( \lambda - \mu -3 -
\alpha (N-3))+2(2 \lambda - 2 \mu -7). $$
Then the common part of the three recurrence relations sets is the
following:
$$c_{s'-k+1}={{(N-s'+k)!} \over {\Gamma\left( (2\lambda -5)/(2\alpha)
-N +s'-k\right)}}[(-2\alpha)^k $$
$${{\Gamma\left( (2\lambda
-5)/(2\alpha) -N +s'\right)} \over {(N-s')! R_{s'kk}}}c_{s'+1}$$
$$+4 \sum_{l=1}^k (-2\alpha)^{l-1} {{\Gamma\left( (2\lambda
-5)/(2\alpha) -N +s'-k+l-1\right)} \over {(N-s'+k-l)! R_{s'kl}}}$$
$$ \sum_{\sigma =1}^{s-s'+k- \tau -l}
\sum_{\tau =0}^1 b_{s'-k+l+ \sigma + \tau N-\sigma+1}(1-
\delta_{\tau 1} \delta_{lk}], {~~~}k=1,...,s'$$
In this relation summation is performed under the condition
$2\sigma \le N+1+k-s'-l- \tau$.
$$ b_{jn'-k}= (-4\alpha)^k {{(N-n'+k+1)!} \over {\Gamma\left(
(\lambda -2)/ \alpha -N +n'-k\right)}} \{
{{\Gamma\left( (\lambda -2)/ \alpha
-N +n'\right)} \over {(N-n'+1)! F_{jk1}}}b_{jn'}$$
$$ -2 \sum_{l=1}^k(n'-l-j) \sum_{\sigma=j+1}^s \sum_{\tau=0}^1
(-4\alpha)^k
{{\Gamma\left( (\lambda -2)/ \alpha
-N +n'-l\right)} \over {(N-n'+l+1)! F_{jkl}}}[$$
$$(2(N-j+1)(\lambda -2 -\alpha(N-1))+n'-l+j)b_{j+1n'-l} $$
$$ -2(n'-l-j) \sum_{\sigma=j+1}^s \sum_{\tau=0}^1 b_{\sigma
n'-l-\sigma + \tau +j}(1-\delta_{\tau 1} \delta_{\sigma j+1})]
\},{~~}k=1,...,n'-j+1$$
In this relation summation is performed under the condition
$2\sigma \le n'-l+1+ \tau+j$.
$$ b_{jj}={{2(N-j+1)(2\lambda -3 -2 \alpha(N-j))} \over
S_j}b_{jj+1}.$$

\centerline{\bf 3. Morse potential}

One can come to a a case of molecular Morse potential by a
limiting procedure $x_j =q, {~} q \rightarrow \infty$.
As a result potential (2а) turns into (2b), and
$$ \lambda - \mu \rightarrow 2A+1 \eqno(9) $$
However it's easier to find wavefunctions directly from (3) and
perform the calculations similar to those in section 2.
Substituting (9) into (5) and (7),
one can derive energy levels in Morse potential.
Taking into account
$p(\beta) = -8A$ recurrence relations for the coefficients
in (6) turn out to be the following
$$c_{s'-k+1}=(N-s'+k)![{{(-2A)^k}  \over {(N-s')! R_{s'kk}}}c_{s'+1}$$
$$+4 \sum_{l=1}^k {{(-2A)^{l-1}}  \over {(N-s'+k-l)! R_{s'kl}}}
 \sum_{\sigma =1}^{s-s'+k- \tau -l}
 b_{s'-k+l+ \sigma + \tau N-\sigma+1}], {~~~}k=1,...,s'$$
In this relation summation is performed under the condition
$2\sigma \le N+1+k-s'-l$.
$$ b_{jn'-k}= (-4A)^k (N-n'+k+1)!  \{
{{b_{jn'}} \over {(N-n'+1)! F_{jk1}}}$$
$$ -2 \sum_{l=1}^k
{{(-4A)^k} \over {(N-n'+l+1)! F_{jkl}}}[
(2A(N-j+1)b_{j+1n'-l} $$
$$ -2(n'-l-j) \sum_{\sigma=j+1}^s  b_{\sigma
n'-l-\sigma +j}] \},{~~}k=1,...,n'-j+1$$
In this relation summation is performed under the condition
$2\sigma \le n'-l+1+j$.
$$b_{s-m N}= {1 \over Q_{s-m}} \{ [(N-s+m+1)(1 -4A)-1]b_{s-m+1N} $$
$$+4(N-s+m) \sum_{k=1}^m
b_{s-m+k N-k} -4A c_{s-m} \},{~~}m=1,...,s-1 $$
In this relation summation is performed under the condition $2k
\le N-s+m $.
$$ b_{jj}={{4A(N-j+1)} \over
S_j}b_{jj+1}.$$
It is sufficiently to perform $ \lambda - \mu \rightarrow 2A+1 $ in
the other relations.

Therefore we have found recurrence relations which
determine cofficients $c_j$ and $b_{jk}$ for arbitrary $N$
for Poschl-Teller and Morse potentials. Wave functions are
determined by means of (6).

Let us stress here that at  $ 0<
\lambda - \mu -2 \alpha (N-1) <6 $ and $1<2A- 2 \alpha (N-1) < 7$
wave functions found together with wave functions for $n=1$,
which were found in [7], give the whole discrete spectrum
of the systems in question.

As an example we have found four particle wave function of one of the
excited states in Morse potential (see Appendix).

The results obtained could be used for testing of various approximate
methods of solving many particle quantum system as well as in
supersymmetrical generalizations of such systems. They could also be
useful for investigations of systems close to integrable.

Let us note in conclusion that wave functions of higher levels
with such values of parameters $A$ and $\alpha$ that $n \ge 3$ can be
found in the same way.

\centerline{\bf  Appendix.}

Here we write down explicit form of the coefficients $c_j$ and
$b_{jk}$ for the 14th exited state for four-particle system in
Morse potential.Parameters $A$  and $\alpha$ satisfy inequality
$1< 2A-6 \alpha <7 $. Discrete spectrum consists of 15
levels: the ground state, 4 levels with $n=1 {~}(p=1)$ and 10
levels with $n=2 (p=2)$.  The energy of the level in question
corresponds to $j_l=\delta_{l3} + \delta_{l4}$ and is equal to
$$ E =-2(4A^2-12 \alpha A -16A +14 \alpha^2 +30 \alpha +19)$$
Suppose
$$F(A,\alpha)=(1-2A)(12A-5\alpha-23)(4A-3\alpha-6)(4A-3\alpha-5)$$
$$T(A,\alpha)=2(24A^3-104A^2+42\alpha A +116A
-3)(4A-3\alpha-5)/F(A,\alpha) $$
$$X(A,\alpha)=8A(4608A^4-5376\alpha A^3 -15720A^3+1035 \alpha^2 A^2$$
$$+9504\alpha A^2 +13144 A^2  - 42\alpha A -116A +3)/ F(A,\alpha)$$
$$Y(A,\alpha)=3(73216A^5-248565 A^4 -85632 \alpha A^4 + 150328\alpha
A^3$$
$$+149893 \alpha^2 A^3 +204520 A^3+ 1888 \alpha A^2 +2285 A^2$$
$$-624A -126 \alpha^2 A -606 \alpha A+ 9 \alpha +
15)/[(8A -6 \alpha -9)F(A,\alpha)]$$
$$ Z(A,\alpha)=64A^2(32A^2-24 \alpha A -36 \alpha
-3)(14A-10\alpha-9)/(8A- 6 \alpha-9)$$
$$N(A,\alpha)={{296A -366\alpha-421} \over {8A-6\alpha-9}} $$
$$M(A,\alpha)={{2A^2[3Z(A,\alpha)-N(A,\alpha)Y(A,\alpha)]}
\over {(5A-6 \alpha -7)(3A -4(\alpha+1))}} $$
Then the coefficients of (6) depend on the potential depth $A$
and pair interaction constant $\alpha$  in the following way
$$c_1=2 \left( {2 \over {3A-4 \alpha -5}}(T(A,\alpha)+{{8A} \over
{2A-1}})-{{3A(A-3 \alpha -6)} \over {(2A-2 \alpha -3)(4A-3 \alpha
-6)^2}} \right)d$$
$$c_2={{6} \over {4A-3 \alpha
-6}}d$$
$$ b_{11}=M(A,\alpha){{4A} \over {3A -4(\alpha+1)}}d $$
$$b_{12}=M(A,\alpha)d $$
$$b_{13}={{-4A} \over {5A -6
\alpha-7}}[Y(A,\alpha)+Z(A,\alpha)]d{~};{~~}b_{14}=Y(A,\alpha)d $$
$$b_{22}={{12A} \over {8A -6
\alpha-9}}X(A,\alpha)d{~};{~~~~~}b_{23}=X(A,\alpha)d $$
$$b_{24}=T(A,\alpha)d{~};{~~~~~}b_{33}={{8A}\over
{2A-1}}d{~};{~~~}b_{34}=d $$
Here   $d$  is an arbitrary constant to be fixed by normilizing
condition.

\vskip 1in
{\bf References}

[1] J.Dittrich, V.I.Inozemtsev. J. Phys.A. 1993. V.20.
P.753.

[2] D.V.Meshcheryakov, V.B.Tverskoy. Moscow
University Bull. Physics, Astronomy. 2000. N1. P.{~~~}65.

[3] F.Calogero. J.Math.Phys. 1971. V.12. P.419.

[4] B.Sutherland. Phys.Rev.A. 1971. V.4. P.2019

[5] M.A.Olshanetsky, A.M.Perelomov. Phys.Rep.
1983.  V.94. P.312

[6] V.I.Inozemtsev, D.V.Meshcheryakov. JINR Rapid
Communications. 1984. N4-84. P.22

[7] V.I.Inozemtsev, D.V.Meshcheryakov. Physica Scripta.
1986.  V.33. P.99

\bye